\documentclass[12pt]{article}
\usepackage{amstex,epsfig}
\def\gammamu{\gamma^{\mu}}
\def\gammanu{\gamma^{\nu}}
\def\g5{\gamma^5}

\def\d4k{{d^4k\over (2\pi)^4}}

\def\q{\rm{q}}
\def\barq{\rm{\bar q}}
\def\qq{<\rm{\bar q q}>}

\newcommand{\beq}{\begin{eqnarray}}
\newcommand{\eeq}{\end{eqnarray}}
\newcommand{\beqno}{\begin{eqnarray*}}
\newcommand{\eeqno}{\end{eqnarray*}}
\def\lsim{\mathrel{\rlap{\lower4pt\hbox{\hskip1pt$\sim$}}
    \raise1pt\hbox{$<$}}}         %less than or approx. symbol
\def\gsim{\mathrel{\rlap{\lower4pt\hbox{\hskip1pt$\sim$}}
    \raise1pt\hbox{$>$}}}         %greater than or approx. symbol

\begin{document}
%\begin{center}
%
\title{Pomeron and the Reggeized Glueball/Sigma}

\author{Leonard S. Kisslinger\\
        Department of Physics,\\
       Carnegie Mellon University, Pittsburgh, PA 15213\\
                    and\\
       Wei-hsing Ma\\
       Institute of High Energy Physics, Academia Sinica\\
       Beijing 100039, P. R. China}
\maketitle
%pacs
\indent
\begin{abstract}
It has long been believed that the Pomeron, which has been successful in 
phenomenological fits to high energy scattering data, is associated with
gluonic exchange. By determining the Regge-nucleon vertex in terms of 
previously determined glueball-quark coupling we show that
the Pomeron might be related to the Regge trajectory defined 
by a light scalar glueball/sigma system and a tensor glueball, 
which involves complicated nonperturbative QCD. We predict a tensor
glueball at 2.8 GeV.
\end{abstract}

\vspace{0.5 in}

\noindent
PACS Indices: 12.40.Gg, 12.38.Lg, 13.85.Dz, 13.60.Le
\newpage
\section{Introduction}
\hspace{.5cm}
The Regge picture\cite{regge} is successful for accounting for high energy elastic 
hadronic scattering and diffractive processes, however, Regge trajectories of established
hadrons are not consistent with high energy data\cite{pcol}.
Early in the attempts to fit experiment with phenomenological Regge models
it was suggested that an additional Regge trajectory that could correspond 
to a particle with vacuum quantum numbers at it lowest energy was needed. In order to
fit the behavior of high energy ($\sqrt{s}$) elastic or diffractive cross sections
the pole position in the J-plane, $\alpha$(s), of the Regge pole that dominates high 
energy scattering should have the property that 
$\alpha_{\rm vac}$(0) $\simeq$ 1.0\cite{cf}. This is the Pomeron. 

There has been a problem in understanding Regge phenomenology and high energy elastic 
and diffractive scattering, since none of the Regge trajectories for t-channel exchange 
associated with known mesons can fit into the Pomeron trajectory with  
$\alpha$(0) $\simeq$ 1.0.
There have been many conjectures about the nature of the Pomeron, but the dynamics 
leading to the Pomeron is still not understood. As was observed by a number of workers 
in this field\cite{pcol} it seems that there is no simple resonance or pole on the 
Pomeron trajectory.

A phenomenological Pomeron exchange model with a vector-type Pomeron-nucleon 
vertex was proposed\cite{lp}
\beq
      V^{P-N} & = & \beta \gammamu {\rm F_1(t)},
\label{vertex}
\eeq
where F$_1$(t) is the isoscalar nucleon form factor and $\beta$ is a parameter,
and has been used in a number of fits to high energy experimental
data. We refer to this as the DL model.
Good fits to pp and p${\rm \bar{p}}$ elastic scattering,  diffractive 
dissociation\cite{dl1} and $\rho$-meson electroproduction\cite{dlpl} have
been obtained with a value of the vertex parameter $\beta \simeq$ 6.0 Gev$^{-1}$.
The trajectory found in this model has an intercept $\alpha$(0) = 1.08 and a slope
$\alpha^\prime $(0) = 0.25 GeV$^{-2}$. 
It is the objective of the present work to attempt to derive this vertex parameter
with nonperturbative QCD by assuming that the Pomeron is associated with a 
glueball/sigma type trajectory and using known properties of glueballs and a 
conjecture of a light scalar glueball/sigma system, discussed below.

In terms of Quantum Chromodynamics (QCD) a hadron that is not
a $\bar{q}q$ meson and has vacuum quantum numbers is a candidate scalar 
glueball, and many theorists have stated that it is expected that the Pomeron
is related in some way to glueball exchange. There are two distinct theoretical
efforts related to Regge theory and the Pomeron: the so-calleed soft and hard
Pomerons. The ``soft'' Pomeron is essentially the original Pomeron, which would
give universal fits to elastic and diffractive processes at high energies.
The program of ``hard'' Pomeron physics is a study of high energy, high momentum
transfer process.  See, e.g., Ref\cite{le} for a description of the main features
of these two programs. In the present article we use the terminology Pomeron for 
the soft Pomeron.

The theoretical ``hard'' Pomeron studies are an extension of the first detailed 
theoretical model for Regge exchange, called the multiperipheral 
model\cite{asf}. Attempts to describe the Pomeron by gluonic exchange processes 
started with
a bag model picture\cite{low} and by an explicit model of two-gluon exchange\cite{nu}.
Phenomenological fits using two-gluon exchange have been carried out\cite{dll}.
Attempts to put this model on sounder theoretical ground using perturbative QCD
have led to an integral equation\cite{bfkl} for the two-gluon ladder exchange.
See Ref.\cite{lr} for a review of attempts to derive the Pomeron from gluonic ladders.
There has also been a suggestion\cite{bj},
based in part on the early two-gluon exchange picture\cite{nu}, that experimentally 
observed rapidity gaps are evidence for a gluonic exchange picture of the Pomeron.

However, for the soft Pomeron (the Pomeron) it is expected that 
nonperturbative QCD is required for a microscopic treatment. Although the 
Pomeron trajectory is higher than all meson trajectories, it is possible that 
a scalar glueball could be on the daughter trajectory of the Pomeron, which is the 
conjecture of the present article. 
There have been many theoretical treatments of glueballs using QCD sum rule
methods\cite{gb1,gb2}. Recently, it was shown that mixed scalar glueballs
and mesons have masses in the 1300-1500 MeV region, which cannot be on the 
trajectory of the Pomeron, however,  there is also
sum rule solution at energy far below the region for scalar mesons\cite{lk1}.
More recently it has been conjectured that the low-energy scalar glueball 
is strongly coupled to the sigma/$\pi\pi$ system which results in a broad,
low-energy scalar resonance\cite{lk2}. It is this coupled-channel resonance,
which we call the Glueball/Sigma that we consider here.
With the assumption that the Pomeron is associated with a
glueball/sigma-Regge trajectory, 
a nonperturbative treatment of the Pomeron-Nucleon vertex,  V$^{\rm P-N}$,
can be carried out using the glueball solutions obtained by QCD sum rules.

In the present note we explore the possibility that the low-lying glueball/sigma 
might be related to the Pomeron trajectory in the sense of a daughter 
trajectory\cite{fw} by finding the glueball-nucleon coupling 
for the diffractive region and comparing it to the phenomenological 
Pomeron-nucleon vertex. The nature of the trajectory is not understood, and
the dynamics of the Glueball/Sigma is modelled by fits to experiment and not
understood dynamically. We review the QCD sum rule work on mixed scalar mesons
and glueballs and the idea of the sigma/glueball in Sec. 2;
and calculate the scalar glueball/sigma coupling to the
nucleon in the next two sections, showing that the coupling is consistent with
the phenomenological pomeron-nucleon coupling. We give our conclusions in Sec. 5.
 
\section{Glueballs, the Glueball/Sigma and the Pomeron}
\hspace{.5cm}
Since the Pomeron is almost certainly mainly composed of glue, it is natural
to consider the relationship between glueballs and the Pomeron. In the present
work we are dealing with the scalar 0$^{++}$ states. If such a glueball is
on a Regge trajectory associated with the Pomeron, it would lie on the
daughter, as we discuss below. In this section we shall briefly review
recent theoretical work on scalar glueballs in order the clarify the
nature of the hypothesized Glueball/Sigma. In the energy regions where
scalar mesons and scalar glueballs are both present it is expected that
they should mix\cite{gb1}. In the QCD sum rule method this leads one to
consider a mixed meson and glueball current,
\beq
\label{1}
         \eta(x) & = &  c_1 G \cdot G +c_2 \bar{q}(x)q(x).
\eeq
The constants c$_1$, c$_2$, which are determined by the sum rules, give the
composition of the lowest state as being mainly glueball or mainly meson. 
Using this form\cite{lk1} it was shown that the purely mesonic solution at
about 1 GeV, which would be associated with the f$_o$(980) does not occur,
but mixed meson/glueball solutions are found that can be interpreted to be
the f$_o$(1370), which turns out to be mainly meson, and the f$_o$(1500),
which is mainly glueball.  Most striking is the result that
far below the region of scalar mesons there is a pure glueball solution
in the region 300-600 MeV. The conjecture that this glueball might be
very strongly coupled to the low energy 0$^{++}$ two-pion system\cite{lk2}
follows from  the experimental analysis of possible
glueballs by the BES group\cite{bes}, which showed that glueballs have a
large branching ratio to the sigma, a resonance seen in the low-energy
scalar, isoscalar $\pi - \pi$ system\cite{zb}. We call this coupled-channel
phenomenon the Glueball/Sigma, and investigate its possible relationship
to the pomeron. Of course, since the Pomeron intercept is about 1, the
Glueball/Sigma cannot be on the pomeron trajectory, but could be on the
daughter\cite{fw} to the pomeron. The f$_o$(1500) and  f$_o$(1700) candidate
glueballs\cite{cb} might be on some gluonic Regge trajectories, but not
on the Pomeron or its daughter trajectories.

   A most important aspect of theory is that if the glueball/sigma conjecture
is correct, that is if the low-energy glueball is a pole that drives the
sigma two-pion resonance with a mass and width of about 400 MeV, then from
the experimental Breit-Wigner resonance the gluon-sigma coupling matrix
element is determined. As we shall now show, from this one can predict the
glueball-nucleon coupling and compare it to the phenomenological Pomeron-
nucleon coupling. We do this in the next two sections.

\section{Scalar Glueball-Nucleon Coupling}
\hspace{.5cm}
First let us look at a very simple for  glueball-nucleon coupling by finding
the coupling of a glueball to a quark. Let us assume that the quark is moving
in an external glueball field, as depicted in Fig. 1.
\begin{figure}
\begin{center}
\epsfig{file=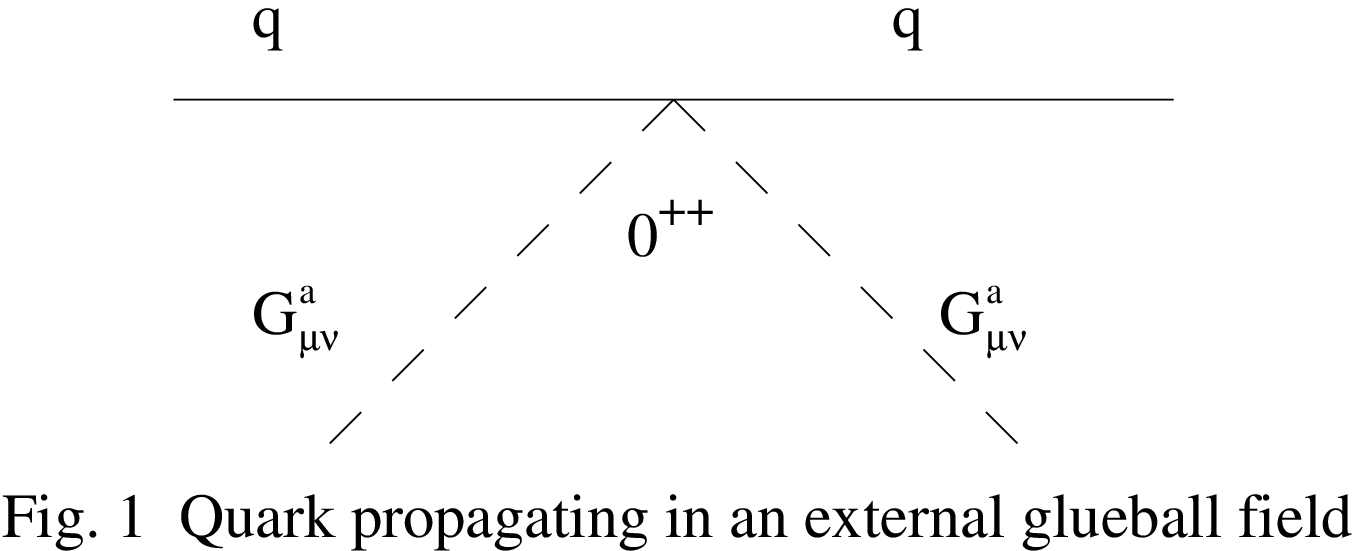,width=10cm}
%caption{}
{\label{Fig.1}}
\end{center}
\end{figure}
This effective quark propagator is given in space-time by the expression

\beq
  S_q(x)^{GB} & = & \int d^4y <T[\barq(x) \q(y){\rm J^{GB}}(y) \barq(y) \q(0)]>,
\label{sgb}
\eeq
where J$^{\rm GB}$, the scalar glueball current, is defined with the
normalization\\
J$_{\rm GB}$ = 3 gG$\cdot$G/(4$\pi$)$^2$. Making use of the
fact that the effective quark propagator defined in Eq.(\ref{sgb}) will
be used in the region x $\rightarrow$ y $\rightarrow$ 0, the integral in
Eq.(\ref{sgb}) can be approximated by the low energy theorem \cite{gb1}
\beq
 \int d^4y <T[{\q}(0) {\barq}(0) \rm{J^{GB}}(y)]> & \simeq & 
 \frac{32}{9} <0|:\bar q(0) q(0):|0>.
\label{let}
\eeq
Using the approximation of Eq.(\ref{let}) in Eq.(\ref{sgb}) we obtain for
the effective guark propagator at momentum k $\rightarrow$ 0
\beq
  S_q(k \rightarrow 0)^{GB} & \simeq & \frac{32 a}{9 (2\pi)^2 m_q \lambda}\ \ ,
\label{sgb1}
\eeq
with a = $\qq (2 \pi)^2 \simeq$ 0.55 GeV$^3$. The quantity $\lambda =
<0|J_{GB}|GB>$ is a normalization factor given in Refs.\cite{gb1,lk1}
Using a quark mass of 8 MeV and a glueball mass of 600 MeV
this gives for the glueball-nucleon vertex
\beq
     V^{GB-N}(t \rightarrow 0) & \simeq & 8.0 GeV^{-1} .
\label{bgb}
\eeq
Comparing to Eq.(\ref{vertex}) the coupling strength is of the order of
magnitude, but is a scalar rather than a vector coupling as in the DL model.

We emphasize that the glueball solution found by using the current ${\rm J^{GB}}$
is not the f$_0$(1500). The  f$_0$(1500) has properties\cite{cb} suggesting that 
it is a mixed glueball/meson, which is consistent with a sum rule solution with a 
mixed ${\rm J^{GB}}$ and $\barq q$ current\cite{lk1} in the 1500 MeV region. It has been
understood for many years that the f$_0$ cannot lie on the Pomeron trajectory\cite{pcol}.
The glueball we assume is part of the coupled low-energy glueball/$\sigma$, which
is observed as a pole at 400 + 400i in the $\pi-\pi$ scalar, isoscalar 
amplitude\cite{lk2,zb}.

\section{High-Enegy Glueball-Nucleon Coupling}
\hspace{.5cm}
In this section we use the methods developed for deep inelastic scattering
to derive the glueball-nucleon coupling for use in high-energy processes,
and see if the result is consistent with the Pomeron coupling.

Recall that inclusive deep inelastic scattering (DIS) cross section can be 
obtained from forward Compton scattering (see e.g. Ref.\cite{am}), $\gamma$ + p 
$\rightarrow$ $\gamma$ + p by using an operator product expansion in a 
light-cone representation. The hadronic tensor for a proton target is
W$_{\mu\nu}$ =  
Im[T$_{\mu\nu}$]/(2$\pi$),
\beq
   T_{\mu\nu} & = &  i \int d^4x e^{ik\cdot x} <p|T[J_\mu(x)J_\nu(0)]|p>,
\label{tmunu}
\eeq
with the electromagnetic current J$_\mu$(x) = $\barq (x)\gammamu Q \q (x)$,
where 
Q is the charge operator. In analogy to DIS, we evaluate T$_{\mu\nu}$ using the
diagram shown in Fig. 2.
\begin{figure}
\begin{center}
\epsfig{file=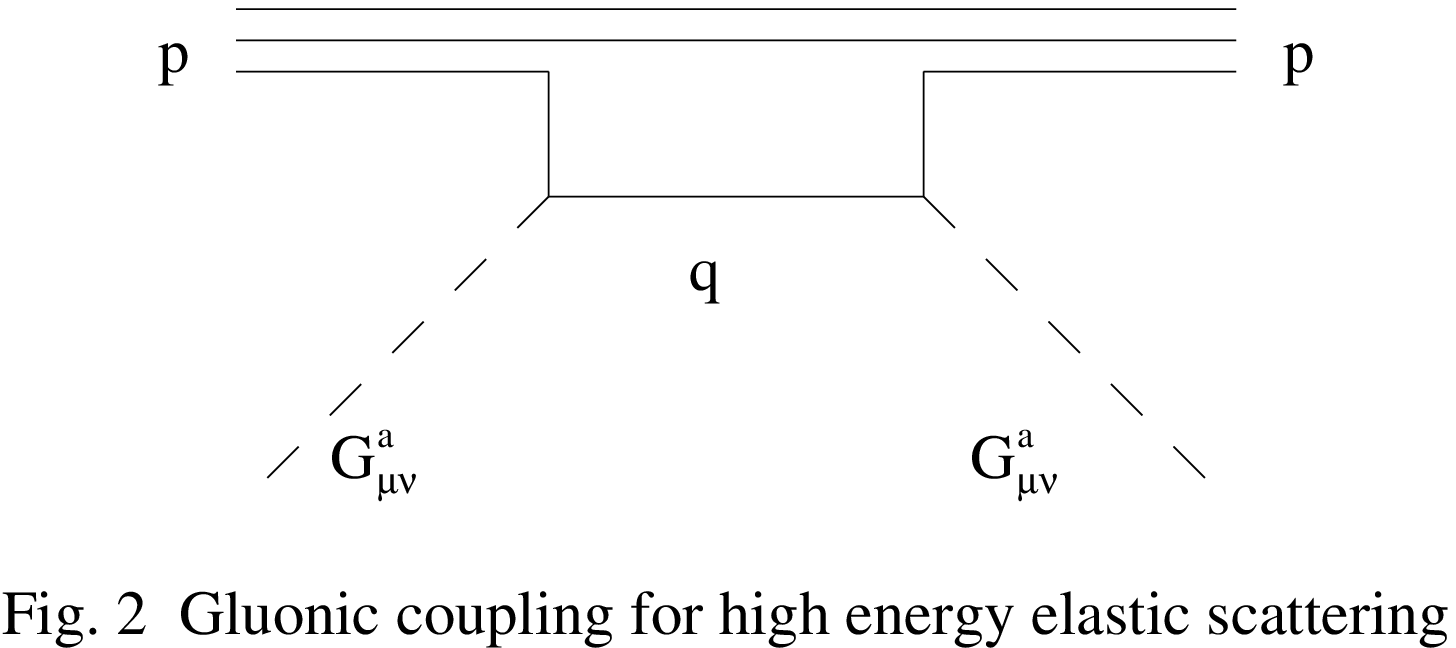,width=10cm}
%caption{}
{\label{Fig.2}}
\end{center}
\end{figure}
Introducing a light-cone representation with momentum
(k$^+$,$\vec{\rm k}^\perp$,k$^-$), one finds in the scaling region
\beq
   W_{+,-} & = &  i\frac{2}{\pi} \int dx^- e^{ik^+ x^-} 
<p|T[q^\dagger_-(x_-)Q^2 q_-(0)]|p>,
\label{wpm}
\eeq
where q$_-$ is a light-cone projection of the quark field. The expression in 
Eq.(\ref{wpm}) gives the parton model for the structure functions, with
W$_{+,-}$ = F$_2$(x)/2x expressed in parton distribution functons, with x
the scaling variable.

Let us consider the forward gluon-proton scattering T-matrix
\beq
   T & = & i \int d^4x d^4y e^{ik\cdot (x-y)} <p|T[J_c(x)J_c(y)]|p>,
\label{t}
\eeq
with the color current J$_c(x)$ = $\barq(x)\gammamu A_\mu(x) \q(x)$.
This is the analog to the forward Compton T-matrix with the electromagnetic
potential replaced by the gluonic color potential.
We use the fixed point gauge, x$_\mu$ A$^\mu$(x) = 0, and 
A$_\mu$(x) = -G$_{\mu\nu}$(0) x$^\nu$/2, with G$_{\mu\nu}$ = $\sum_{a=1}^{8}
\tau^a$ G$^a_{\mu\nu}$, $\tau^a$ being the SU(3) color operator.
Keeping the lowest-dimension contractions one finds the standard form
\beq
  T & = & \frac{i}{4} \frac{\partial}{\partial k^\alpha}
\frac{\partial}{\partial k^\beta}  i \int d^4x d^4y e^{ik\cdot (x-y)}
Tr[S_q(-x) \gammamu S_q(x) \gammanu <g^2 {\rm G_{\mu\alpha}} G_{\nu\beta}>]
\label{tt}
\eeq
Proceding as in DIS, and making use $\partial S_q(k)/\partial k \rightarrow
S_q/m_q$ in the limit k $\rightarrow$ 0, one finds
\beq
       T & \simeq & K \frac{\partial}{\partial k^+} \int dx e^{ikx} S_q(x)\\
\nonumber
       K & = &  \frac{2 a}{27 m_q \lambda}.
\label{k}
\eeq
Taking the value 8 MeV for the quark mass we find for the vertex parameter in
$V^{GB-N}(t\rightarrow0)$
\beq
     \beta & \simeq & 6.6 GeV^{-1},
\label{bgb2}
\eeq
compared to the phenomenological DL value\cite{dl1} of 6.0 GeV$^{-1}$.
From this we conclude that the Pomeron trajectory could be closely related 
to the coupled scalar glueball/sigma system. 

Of course the value of  $\alpha$(0) $\simeq$ 1.0. is
not consistent with a simple interpretation of a light scalar glueball on the Pomeron
trajectory, as has been known for decades\cite{pcol}. However, one possible
interpretation is that the Glueball/Sigma system is on the
first ``daughter'' trajectory\cite{fw} of the Pomeron. Since daughter trajectories 
satisfy $\alpha$(0)$_{\rm daughter}$ =$\alpha$(0) - 1.0, and the daughter and Regge 
trajectories are parallel, this is a possible solution. In potential models\cite{fw}
it is shown that the residue of the daughter results in a cancellation of an unwanted
singulaity in the Regge picture. Daughters also appear in dual models\cite{pcol}.

A recent calculation\cite{lm}
has shown that the $\xi$(2230) observed at BES\cite{bes}, which has some 
characteristics of 
a glueball, but is not yet established as a 2$^{++}$ resonance, might be on the
Pomeron trajectory. Using experimental pp cross sections the
branching ratio to p${\rm \bar p}$ was predicted, which serves as a test of the
nature of the $\xi$(2230). From the phenomenological slope of the 
Pomeron/daughter trajectory\cite{dll}, $\alpha^\prime $(0) = 0.25 GeV$^{-2}$,
we predict that
a tensor glueball on the same daughter trajectory as the light glueball/sigma
will be found at about 2.8 GeV, so that a tensor glueball associated
with the scalar Glueball/Sigma is predicted to occur at an energy 
about 600 Mev higher than the $\xi$(2230). As noted in Sec. 2, the f$_o$(1500)
and f$_o$(1710) candidate scalar glueballs are not on the Pomeron or daughter
trajectories.

\section{Conclusions}
\hspace{.5cm}
We have shown that the low-energy glueball that recently has been proposed as a coupled
glueball-sigma system couples to a nucleon with the strength roughly in agreement with
phenomenological Regge Pomeron fits to a number of scattering and production 
experiments at high energy. In this picture we propose that a light scalar 
Glueball/Sigma lies on a
daughter of the Pomeron Regge trajectory. This is consistent with the  $\xi$(2230)
being a tensor meson/glueball on the Pomeron trajectory and a higher-mass tensor
glueball, we predict to be at 2.8 GeV, being on the daughter trajectory with the 
scalar glueball.
The glueball/sigma system is complicated and 
the dynamics of this system are certainly not understood by the authors, however, 
this could also be the nature of the Pomeron.

The authors would like to acknowledge helpful discussions with Dr. L-C. Liu.

The work was supported in part by the National Science Foundation
grants PHY-9722143 and INT-9514190.

\end{document}